\documentclass[12pt]{article}
\input{epsf.tex}
\begin{document}

{Partial level density of the $n$-quasiparticle excitations in the nuclei
of the $40 \leq A \leq 200$ region}\\\\
 {A.M. Sukhovoj, V.A. Khitrov }\\\\

 {\it Frank Laboratory of Neutron Physics, Joint Institute for Nuclear
Research \\ 141980 Dubna, Russia}\\\\

Level density and radiative strength functions
are obtained from the analysis of two-step cascades
intensities following the thermal neutrons capture.
The data on level density are approximated by the sum of
the partial level densities corresponding to $n$ quasi-particles excitation.
The most probable values of the collective enhancement factor of the level
density are found together with the thresholds of the next Cooper nucleons
pair breaking. These data allow one to calculate the level density of
practically any nucleus in given spin window in the framework of model
concepts, taking into account all known nuclear excitation types.
The presence of an approximation results discrepancy with theoretical 
statements specifies the necessity of rather essentially developing
the level density models. It also indicates the possibilities to
obtain the essentially new information on nucleon correlation
functions of the excited nucleus from the experiment.

\section{Introduction}
First of all, obtaining data on the level density $\rho$ and radiative strength
functions $k$ emission of nuclear reaction products is a way to universally
test nuclear models. Their accuracy is quite unequivocally limited by the
experimental data quality used for both creation of new modeling representation,
and for its parameterization. Naturally, the experimental data used for this
purpose should be the most reliable. At present, for evaluating of [1] neutron
cross sections it is required that there is a high reliability of modeling
representations on the level density in a wide nucleus excitation energy
interval. It is also important for describing [2] their fission process.
That means they are needed for quite practical applications of nuclear physics.
Thus, materials' and nuclear power cycle fuel's profitability and safety directly
depend on the accuracy of their cross sections evaluation.

The history of developing the techniques for experimental determination of
$\rho$ covers several decades, but the significant part of the experimental
data obtained up to now has too large systematical error, which is
essentially unavoidable within the framework of existing experimental analysis
algorithms. It is connected to extraction peculiarity of the level density and
radiative strength functions from experimentally measured spectra.
This regards the situation, when the insufficient resolution of nuclear
reaction's
products, provided by the existing spectrometers, does not allow one to
resolve the individual levels of a nucleus. In wide energy intervals, where the
nuclear reactions' products are registered, the intensity of spectra
simultaneously depends on both $\rho$ and $k$. Errors transfer results in
a large increase of systematical errors in the found values. It always occurs
when there are systematical experimental errors of the measured spectra and the
strong correlation of parameters.

Currently, there is a set of various advanced modeling representations [3] for
predicting both the level density and the radiative strength functions of the
primary gamma-transitions following compound states decay.
However, the accuracy of $k$ values, calculated with their help, is unknown for
the low-energy primary gamma-transitions following neutron resonance decay.
Therefore, contemporary techniques, in which parameters of a nucleus are
extracted from the measured spectra of the nuclear reactions gamma-rays, can
provide no high accuracy for either level density or the radiative strength
functions. It is impossible if the modeling representations on one of these
values are used to determine the other.

These circumstances demand an obligatory revealing of all of the most essential
systematical  errors of $\rho$ and $k$ certain values by comparing any model
representation with the experiment.

\section{Systematical errors of the level density and radiative strength
functions in the experiments of various types}

By present time four different techniques to determine functional dependence
of the level density are realized. They are based on the following experimental
data:

1. Nuclear evaporation spectra in reactions of various types;

2. Full spectra of gamma-transitions for various energies of the decaying levels;

3. Intensities of the two-step cascades appearing at the thermal neutrons
capture;

4. Distribution of the random intensity values of individual energetically
resolved two-step cascades with the energy of their intermediate level
$E_i <0.5B_n$.

Here, it is necessary to add a completely independent technique of [4]
to determine the deviation sign of the experimental level density from any
model prediction. The corresponding analysis uses the data on random
fluctuations of the intensity ratios of secondary gamma-transitions at the
resonance neutrons capture. Its results obviously contradict the conclusions of
the 1-st and 2-nd techniques and qualitatively confirm the basic of the 3-rd
and 4-th methods that the level density is smaller than that given by the
present standard models.

As it is stated above, all specified techniques always result in $\rho$ and $k$
values having unknown systematical errors. The most exact of their estimation
can be obtained only by comparing their values, determined in various and
complementary independent experiments. The comparison quite uniquely shows that
the level density in the region of the half of neutron binding energy,
found within the framework of the first two
techniques is $\sim 3-5$ times (in a number of nuclei - and more)
larger than that determined according to the 3-rd and
4-th algorithms. At lower and higher excitation energies this discrepancy
decreases because densities of resolved low-lying levels and of neutron
resonances are always normalized to the same values in techniques 1-3.

Thus, during development and parameterization of $\rho$ models, results of
application of the level density, obtained by using of the listed
above techniques, may bring to rather fallacious ideas about this parameter of
a nucleus.

1. Practically, all theoretical models of level density [3] are completely
developed on the base of the experimental data on $\rho$, obtained similarly
to [5] from evaporation spectra of nucleons or light nuclei in various nuclear
reactions. In order to determine the experimental level density, the model-set
values of transmission coefficient $T$ of a nucleus surface have been used for
the emitted nucleons or light nuclei. Up to the present the analytical method
of an optical model of a nucleus is used for their determination, providing
that there is no determinate choice of nuclear potential used for calculation.
It is necessary to take into account, that 283 various parameterizations of
optical potential were developed to describe the neutron-nucleus interaction
process; 101 - for proton, correspondingly [3]. Such a quantity of entrance
factors needed for calculation of the transmission coefficient in reaction
(n,p) [5] serves as a serious reason for assuming that there are very big
errors in extraction of level density by means of the first technique.
Naturally, determination of level density with the minimal error requires one
to calculate $T$ values within the framework of nuclear models with the
guaranteed accuracy which exceeds that needed for $\rho$ determination.
The latter, by magnitude scale, may be equaled to the modern accuracy of
observation of the spectrum intensity. In the worst case it makes some tens of
percents. Currently, such an accuracy of $T$ calculation is most likely
unattainable, even if not only the optical model of a nucleus, but also more
modern nuclear models are at use. As a bright example, it is possible to
indicate the comparison of calculated and experimental values of strength
functions of (d,p) or (d,t) type reactions. Now it is fulfilled for many nuclei
in experiments with high resolution at Tandem Accelerator of the University and
Technical University of Munich up to the approximately half of neutron binding
energy. In all cases, the significant information on the excited levels
structure, which allowed one to make a comparison with the nuclear theory with
the excitation energy of up to several MeV, is obtained through these
experiments. Such techniques as [5] do not take this fact into account.

The comparison of (see, for example, [6]) the results obtained in Munich with
calculations within modern nuclear models shows that details of a fragmentation
process of any states of nuclear potential over real levels with excitation
energy up to several MeV can be reproduced only with a significant error.
First of all, this circumstance is caused by an insufficient accuracy of the
notions incorporated in the modern model on the nature of the nucleons
interaction in a nucleus, as well as by the inevitable approximations of its
account in anyone of them. The uncertainty of parameterization of the concrete
nuclear models (optical, quasiparticle-phonon, models of interacting bosons and
fermions, etc) always brings an additional error to the process of determining
[5] $\rho$ with the help of calculated $T$ values. First of all, it is pointed
at by the unavoidable and rather significant discrepancy between the
experimental and model-calculated values of energies even for the most
low-lying levels of the simplest structure. Moreover, it is also emphasized by
the essentially greater one at the calculation of the nucleus parameters in
case when excitation energy is higher than 1 - 3 MeV.

Discrepancy between the notions of the level density extraction from the nuclear
evaporation spectra and the technique of determining the excited levels
structure by the levels excitation probability in nuclear reactions is a
principal one. Basic thesis of a technique [6] follows from the experiment and
cannot cause doubts. Reasoning from this, the existence of big systematical
errors in $T$ calculation is expected in case of using the theoretical models
for the experimental determination of the level density [5] from the
evaporation spectra.

One should not exclude a possibility of another explanation of a discrepancy of
$\rho$ values determined by the first and the third techniques. For example, it
may occur if a discrepancy of model notions and the experiment for spin
dependence of the level density is not taken into account at energies lower
than several MeV. Particularly, as there is a possible, but theoretically
unconsidered, strong dependence of collective enhancement factor in level
density [3] on a level's spin.
If such an appreciable effect exists, it is displayed for spins J exceeding
$\sim 6$.
It follows from maximal spin values of neutron resonances in $^{150}$Sm,
$^{177}$Lu
and from $\rho$ value obtained for these and other nuclei.

2. Very serious and, probably, unsolvable problems arise at determination [7,8]
of $\rho$ and $k$ parameters of the cascade gamma-decay process of the nucleus
from the gamma-rays spectra $S$ of any nuclear reactions, in which the
nucleus-product is excited up to the energy of 5-8 MeV and higher.

Rather simple modeling [9] of process of transferring errors $\delta S$
of the determination of the total gamma-spectra intensity into the ones of
primary gamma-transitions [7], subsequently extracting the values of the level
density and radiative strength functions from them, has shown that the value of
$\delta \rho /\rho \leq 40-50\%$  may be obtained within the framework of
[8] type procedure. But it is possible only under the condition of all
relative systematical errors of the measured gamma-spectra belonging to the
region $\delta S/S <0.001-0.003$ for all energies of emitted gamma-transitions
and depopulated levels. This conclusion is just, at least, for the case of
monotonous increasing or decreasing of the systematical errors $\delta S$
as the nucleus excitation energy changes. Such an accuracy of measuring the
intensity of total gamma-spectra by means of scintillation detectors is most
likely unattainable for the present. It appears so, both because of the
instability of their work, and of the impossibility of such an exact subtraction
of Compton background from the instrumental spectra.
Besides, in case when there is a nucleus with the high level density, the
subtraction errors are the greatest.

Additional and very essential systematical errors in the data for $\rho$
and $k$, obtained by means of a technique of the primary gamma-transitions
spectra determination [7] in corresponding experiments, are conditioned by its
authors applying their own [8] algorithm for searching the maximum of likelihood
function. It is used instead of the updates of Gauss method developed for
completely similar cases and well tested by mathematicians.
It is impossible to estimate the systematical errors related to it, as the
method [8] can provide no information on position and value of its false maxima
unlike the modern variants of Gauss method.

The assumption stated in [8] deals with the equality of radiative strength
functions of gamma-transitions of equal multipolarity and energy for
depopulating levels with different energies. It contradicts the basic ideas of
modern nuclear models (see, for example, [10]) and results in the unknown
systematical errors occurring both in $\rho$ and in $k$. 


Therefore, there is a necessity to develop the essentially new techniques for
$\rho$ and $k$ determination. These should provide their extremely possible
reliability. As mentioned above, such techniques cannot be created on the basis
of any of the model notions for obtaining the experimental values of level
density and radiative strength functions used both separately, and combined.
That is why any of the newly developed techniques should be both
model-independent, and determine the level density and radiative strength
functions of the reaction's emission products. It should be simultaneously
accompanied with providing the data accuracy of the modern experiment for the
whole of the excitation energies range of the studied nucleus from the most
suitable, for this purpose, reaction products spectra. This operation should be
realized within the framework of the insufficient opportunities (for the
guaranteed problem solution) and of essential limitations of a mathematical
statistics algorithm developed for such purposes. Exactly at this stage of
determination of the investigated phenomenon parameters from strong correlating
experimental data there appear the most serious systematical errors for both
level density and radiative strength functions.
It should be noted that in the mathematical expressions, describing the
experimental spectra, the nonlinearity of the sought parameters connection is
exactly responsible for the significant reduction of the area of pairs'
parameters values with a hundred-percent correlation concerning a case of the
linear equation systems. On this account, the degenerate systems of the
nonlinear equations can have infinite number of solutions, all of which are in
the limited intervals of their values.

\section{The potential of modern experiment to precisely determine the main
parameters of a nucleus}

On one hand, proceeding from the base principles of mathematical statistics,
the extraction of a number of unknown parameters from the data of an experiment 
demands a superfluous number of values to be determined in its course.
On the other hand, the system of the sought parameters functional connections
with the measured values has to provide both unambiguity of determination and
minimum of errors of the sought values. Most likely, these two conditions cannot
be satisfied in cases of obtaining the level density $\rho$ from evaporation
spectra [5] of any nuclear reactions or the radiative strength functions $k$
both from total spectra [7] of gamma-rays following neutron radiative capture
and from nuclear reactions at the charged particles beams [8].

Therefore, the experimental data of other type are necessary in order to get
main nuclear parameters with higher accuracy than achieved at the present.
The analysis [11] of the experimental data on cascade gamma-decay of one
(or several) of the highly excited levels (compound states), fixed at energy
$B_n$, satisfies the listed above requirements to the maximal degree.
First of all, these are the two-step cascade intensities
\begin{equation}
 I_{\gamma\gamma}(E_1)=\sum_{\lambda ,f}\sum_{i}\frac{\Gamma_{\lambda i}}
{\Gamma_{\lambda}}\frac{\Gamma_{if}}{\Gamma_i}=\sum_{\lambda ,f} 
\frac{\Gamma_{\lambda
i}}{<\Gamma_{\lambda i}> m_{\lambda i}} n_{\lambda 
i}\frac{\Gamma_{if}}{<\Gamma_{if}> m_{if}}
\end{equation} 
following the thermal neutron radiative capture, connecting a compound state
$\lambda$ and a group of low-lying levels $f$, in function [12] of energies $E_1$
of their primary gamma-transition. Nucleus excitation energy (energy of the
intermediate level of the cascade $E_i$) is unambiguously determined by energy
$E_1$:  $E_i=B_n-E_1$. Cases of full absorption of the two-step cascade
gamma-transitions energy, necessary for the experimental definition of
$I_{\gamma\gamma}$,
are concentrated in the narrow peaks of the sum coincidence spectrum (Fig. 1)
obtained with the help of ordinary HPGe-detectors. In the same experiment an
inevitable background is measured with the maximal possible accuracy.

Functional (1) depends both on ratio of partial and total radiative widths
$\Gamma $
of primary $E_1$ and secondary $E_2$ gamma-transitions of cascades between
levels $\lambda$, $i$ and 
$f$ and on number of levels $n(m)=\rho \times \Delta E$ excited in different
energy intervals $\Delta E$.
A degree of detailed elaboration of finding the form of energy dependence
the level density and radiative strength functions is determined by the optimum 
interval's width of an averaging-out of cascades intensities.
Technically, its limiting value equals the HPGe-detectors resolution.
However, combined with the inevitable partial width fluctuations, real
capabilities of the modern detector and computer facilities limit $\Delta E$
interval width, over which nucleus excitation energy of 50 keV value
(or a little larger) is distributed.

Cascade transition type (dipole electric or magnetic) and an excited
intermediate level's $i$ spins and parity are unequivocally determined by known
$J^{\pi}$  values of $\lambda$ and $f$ levels. Practical absence of cascades
between levels with $|J_\lambda - J_f |>2$  excludes the necessity of
accounting for transitions with higher multipolarities in analysis like [11].
For convenience of direct comparison of the obtained radiative strength
functions with E1-, and M1-transitions in nuclei with different mass $A$, it is
appropriate to determine them in the following form:
\begin{equation} k=\Gamma_{\lambda i}/(E_{\gamma}^3\times
A^{2/3}\times D_{\lambda})
\end{equation}
and to use their ratios $k(M1)/k(E1)$ for mutual normalization of data,
experimentally measured near $B_n$. 

Intensity of cascades in expression (1) is proportional to a derivative
$dk/dE$  and, as a first approximation, it is inversely proportional to $\rho$.
It provides the maximal sensitivity of the experiment in the range of $\rho$
lowest values. So, in the field of excitation energy, the influence of the
nuclear structure on parameters of studied nuclear reaction should be maximal.
The essentially new connection type between the sought parameters, compared with
the ordinary evaporation and gamma-spectra, provides smaller influence of
correlation parameters on their real error.

The specified experiment has two sources of the ordinary, but potentially rather
significant, systematical errors. They are connected to a possibility:

(a) Full compensation of a divergence between the experimental and calculated
intensity values of the cascades with primary transitions $E_1$ by a divergence
of an intensity opposite sign of cascades with secondary transitions from the
same energy intervals and

(b) Presence of an excessive systematical error at the normalization of
$I_{\gamma\gamma}$ intensities.

The first type error is related to the impossibility of instrument determination
of the quanta ordering in the bulk of observed cascades. To full extent it is
visible in analysis [13], using for extraction of the information on $\rho$
and/or $k$ the fact that the intensities directly observed in the experiment and
calculated for different tested functional dependences of parameters are equal
to their values for all cascade quanta energies. 
According to [9] this error leads to that the experimental spectra can be
precisely reproduced by the level density and the radiation strength functions
distinguished from the sought ones by some tens of times. The indicated error
is practically reduced to zero only in the case, when the intensity of cascades
in function of their primary transition energy is determined from the
experimental spectra [12]. The corresponding algorithm uses the experimentally
proved fact: even in nuclei with the maximal level density there are pairs of
energetically resolved peaks, corresponding to the cascades with $E_i<3-5$ MeV,
which concentrate the main part of sum cascades intensity. A threshold of their
intensity registration is much less than the average value even when rather ordinary
detectors are used. Cascades with primary gamma-transitions from the same energy
interval are registered mainly in the form of continuous distribution of small
amplitude. Total background in any experimental spectra of two-step cascades
intensities is equal to a determination error of a substrate under peaks of
full capture of the cascade energy by the HPGe-detectors (Fig. 1). Practically,
very exact algorithm, realized in [14] in combination with the results obtained
by nuclear spectroscopy, allows one to determine the quanta ordering with high
accuracy. Consequently, it makes possible for one to find the form of energy
dependences of experimental values $I_{\gamma\gamma}(E_1)$ with a relative
systematical error less than [15] 0.01 - 0.05.

The approximation of distribution of the resolved cascades intensity random
values in small energy intervals of their intermediate levels allows [15] both
determination of the corresponding error $\delta I_{\gamma\gamma}$, and direct
estimation of the most possible level density in any energy intervals of
cascade intermediate levels below $\approx 0.5B_n$.
It happens within the framework of a hypothesis about Porter-Thomas's law on
fluctuations of the primary transitions intensities with the subsequent
extrapolation of the approximated distribution below gamma-quantum registration
threshold. In all of the analyzed nuclei the level density either practically
coincides with the data [11], or differs from the latter much less, than that
measured with techniques [5] and [8].

Practically, expected systematical error $\delta I_{\gamma\gamma}$ of the
intensity cascades used in a technique [11] is almost fully determined by an
error of measuring of the most intense primary gamma-transitions of cascades
with low-lying intermediate levels. Comparison of the corresponding data from
[16] and [17] shows that its value may be estimated as $\approx 10-20\%$.
Coefficients of an error $\delta I_{\gamma\gamma}$ transfer onto the values
$\delta \rho$ and $\delta k$ are estimated in [18] for even-odd isotopes of
tungsten and osmium. It was made for all of their excitation energy values.
An error variation in an interval of
$-25\% \leq \delta I_{\gamma\gamma}/I_{\gamma\gamma} \leq 25\%$
changes values, found for the level density below $0.5B_n$, less than by 2 times,
and for radiative strength functions - less than by 3 times. 

It should be noted that the error in determining of $I_{\gamma\gamma}$ can be
larger than $\pm 25\%$ in the nuclei where the thermal neutron capture cross
section was measured with the big relative uncertainty. In these cases, the
amplitude of the expected relative errors of level density and radiative
strength functions can significantly increase as compared with the results
obtained in [18]. For some nuclei [6], this problem can be completely removed
by comparison of the measured gamma-transition intensities with the known
absolute intensities of gamma-rays related to the following $\beta$-decay of the
compound nucleus. One can only assume, however, that the relative errors of the
measured capture cross sections for the most nuclei represented below are, in
average, close to zero and have different sign.

The third systematical error, by decreasing of the importance, is introduced
into the results of technique [11] by using the notions of the identical form
of energy dependences of radiation widths of the primary and secondary
gamma-transitions of given multipolarity and energy. For the first time, the
problem was formulated in the obvious terms and partly solved [19] at the
experimental $\rho$ and $k$ determination from intensities $I_{\gamma\gamma}$
of the two-step cascades. Its partial solution is achieved by using the
experimental information on a full cascade population of the maximal number of
levels in the bottom part of the investigated excitation energies region.
The main result of the joint analysis [20] states that the level density
decreases at the account of radiative strength function dependence on the energy
of depopulating level as compared with the data [11]. It regards the two-step
cascades intensities distribution onto low-lying levels and the data on
population the high-lying ones for two tens of nuclei.
 
In a technique [11] the examined systematical error displays itself to much
smaller degree than in [8]. Due to that the revealed in radiative strength
functions of secondary gamma-transitions changes have a different sign for
various excitation energies, they are summarized in $\Gamma_i$ with a
multiplier $E_{\gamma}^3$ (2) and, consequently, are relatively less in
changing the total radiative width of cascades intermediate level.
This conclusion is just only for the obtained by the present time data on
intensities of two-step cascades to final levels with energy $E_f < 0.5-1$ MeV.

As in any other experiment, at the extraction [11,20] of the cascade gamma-decay
parameters, the existing ideas are used as well as constants and earlier
established concrete nucleus's parameters. First of all, it concerns:

- the number of low-lying levels determined by nuclear spectroscopy [21],

- an average spacing $D_{\lambda}$ between the neutron resonances, $\Gamma_{\lambda}$
total radiative width [22] of a decaying compound state.

The corresponding data bring also additional, but correlating for various
techniques, errors into the found $\rho$ and $k$ values. They can be significant
only if the existing ideas of a nucleus contain essential errors. For example,
it is so when the probability of excitation by a neutron of a level with a
given $J^{\pi}$ higher than neutron binding energy, strongly depends on its
structure. In such a hypothetical case, value of $D_{\lambda}$ can be
essentially overestimated. It may be also possible that the branching
coefficients at the decay of even rather long-living levels noticeably depend
on their excitation way.

If not to take into account such exotic opportunities, the information on
cascades of gamma-quanta allows one to determine $\rho$ and $k$ without using
the model-calculated values with the smallest possible systematical error for
any stable target nucleus at any neutron beam.
It is also needed to carry out [23] an independent check of the found values of
the level density and strength functions by comparing the calculated and
experimental total gamma-spectra. It can be fulfilled at capture of both
thermal and, in principle, resonance neutrons.

Thus, comparison of possible systematical errors of $\rho$ and $k$ values in
experiments of various types shows a basic advantage of techniques [11,20] over
the known ways of determining only the level density [5] or of simultaneous
determining the same value together with absolute ones of radiation strength
functions [8]. It is caused by the following:

(a) due to the instrument selecting only cases of full capture of the cascade
energy, high resolution and stability of semi-conductor detectors the two-step
cascade spectra are measured with a smaller systematical error as compared with
the gamma-spectra, used by technique [8];

(b) the value $I_{\gamma\gamma}$ depends on absolute value of the level density
(used by the alternative techniques in their existing [5,8] variants, spectra
do not depend on the absolute $\rho$ and $k$ values);

(c) the transfer coefficients of a spectrum's error onto the 
$\delta \rho$ and $\delta k$ values in method [11,20] is   $\sim 2$ orders 
smaller than [8];

(d) owing to the condition (b) there is a physically determined and essential
limitation of the interval of possible $\rho$ and $k$ values, precisely
reproducing both the intensity of cascades in energy functions of their
primary transition, and other functionals of cascade gamma-decay process of
any compound nucleus.

(e) especially, it is necessary to emphasize that fixing of spins levels, for
which the level density is determined, is practically unequivocal in expression
(1) and there were no similar experimental results on this parameter till now.

It should be noted that there is an essentially unavoidable error of $\rho$ and
$k$ extraction from cascade intensities. It is caused by an excess of the
unknown parameters number in (1) over the number of the experimentally measured
values $I_{\gamma\gamma}$. Accordingly, the concrete values $I_{\gamma\gamma}$
can be reproduced with an equal $\chi^2$ by infinite set of parameters
determining them. But the region of possible $\rho$ and $k$ values is always
limited, providing that the relation of strength functions of primary and
secondary transitions of equal energy and multipolarity is unequivocally fixed
for any energies of decaying levels. Its width is minimal in the case of using
all of the available experimental information on the investigated nucleus for
determining the cascade gamma-decay parameters.
At following these conditions, an interval of the most probable $\rho$ and $k$
variation does not in reality exceed several tens of percents if the value
$I_{\gamma\gamma}$, used in [11,20], is [12] function of energy of the only
primary cascade transition.

Potentially, obtained $\rho$ and $k$ values can be distorted because of
the influence of all three cascade levels structure on their intensity.
The structure of a decaying compound state can influence $I_{\gamma\gamma}$
in a wide excitation energy interval, intermediate - locally.
The influence degree of the structure of two-step cascade final level on its
intensity is evidently shown in [19,20].
Corresponding effects for cascades primary gamma-transitions can be estimated
and reduced only at their being studied in many neutron resonances. At present,
a variation of the level density in concrete nucleus with respect to its average
general trend (Fig. 2-8) can be accepted as the top value of an influence effect
of resonance structure on the determined $\rho$ and $k$ values. 

\section{On a basic possibility of the precise model-description of the modern
data on the experimental level density}

The general form of dependence of the most reliable modern values of the level
density on the investigated nucleus excitation energy points at [11,20] the
presence of at least two ``step-like" structures below neutron binding energy
with its faster increase in between than it is predicted by a notion of a
nucleus, as a system of non-interacting Fermi-particles.
It means that, at least, at two excitation energies the abrupt change in the
excited levels wave functions structure occurs in a nucleus. The unique factor,
known by the present to be capable of providing such a change is the breaking of
nucleons Cooper pairs with addition of two quasiparticles to the existing ones,
as well as fast increasing the level density at increasing excitation energy.

In modern theoretical notions [3], the level density at a given nuclear
excitation energy $U$, spin $J$ and parity $\pi$  is expressed through the
density $\rho_{qp}$ only of the quasiparticle excitations and its vibration
and rotational (for the deformed nuclei) enhancement coefficients $K_{vibr}$
and $K_{rot}$, respectively: 
\begin{equation} \rho(U,J,\pi)=\rho_{qp}(U,J,\pi)
K_{vibr}(U,J,\pi) K_{rot}(U,J,\pi) = \rho_{qp}(U,J,\pi) K_{coll}(U,J,\pi).
\end{equation}
 
For the further analysis of the experimental data it is expedient to unite
coefficients of vibration and rotational increase in the level density in the
general coefficient of its collective enhancement $K_{coll}$. In the level
density from the analysis [11,20] the basic contribution to its value is brought
by the effect of vibration. In an examined case the effect of rotational
enhancement of the level density for the deformed compound even-odd nucleus is
less than the experimental data error. Probably, it is a little more for nuclei
with neutron resonance spins of  $J \geq 2$. By the order of magnitude in a
neutron binding energy range, it is expected that $K_{coll}$ value for the
complete level density is [3] in an interval: $10 < K_{coll} < 100$.
There is no experimental information on dependence of $K_{vibr}$
on $U$, $J$, and $\pi$.
Modern theoretical ideas of this account admit a significant change in $K_{vibr}$
when changes $U$ up to the change of [24] its functional forms dependences on
nucleus excitation energy. The presence [11,20] of rather reliable experimental
data for the sums of radiation strength functions of dipole cascade transitions
allows, basically, to solve this problem by creation of precise models of
strength functions. This opportunity is caused (see, for example, [10]) by
known distinctions in values of partial radiation width from a ratio of
quasiparticle and vibration components in the structure of excited (decaying)
levels. However, currently, there are no theoretical models of such level [3].

Therefore, the further analysis of the level density is possible to be carried
out only within the framework of zero assumptions of $K_{coll}$ independence on
the nucleus excitation energy in the interval of $\sim 0.5-3$ MeV up to $B_n$.
For the first time, it allows one to receive direct experimental information on
partial density of quasiparticle excitations with various number of
quasiparticles in the specified excitation energy intervals for some nuclei with
their masses of $40 \leq A \leq 200$.

The possibility of determining the partial level density $\rho_{n}$
with a given number $n$ of the excited quasiparticles for $U$ nucleus energy
excitation
\begin{equation}
\rho_{n}=\frac{(J+1)exp(-(j+1/2)^2/(2\sigma^2))}{2\sqrt{(2\pi)}\sigma^3}
\frac{g^n(U-E_n)^{n-1}}{((n/2)!)^2(n-1)!}  \end{equation} 
has been found by Strutinsky [25].
For the first time, he has obtained simple functional dependence of the nucleus
excited states density (the second coefficient in (4)).
In more general form than expression (4), the model takes into account the
existence of proton and neutron Cooper pairs (in the first turn [2] - for
fissionable nuclei). All kinds of its modern
notions are included in reviews [3], being quoted from original publications.
Modern state of methods of partial density calculation is analyzed by
B\v{e}t\'ak and Hodgson in [26].

Practically, for comparison with the experiment [11,20] within the framework of
existing [3] theoretical notions of a nucleus, it is necessary to choose value
of the spin cutoff factor $\sigma=f(n,U)$ for the given Cooper pair and
excitation energy, together with the energy $U-E_n$ of the excited
quasiparticles. Density of single-particle levels $g$ for the presented here
nuclei is known from the data on neutron resonances.
The parameter $g$ can take into account shell
inhomogeneities of single-particle spectrum and depend on $U$.
These rather theoretical opportunities have not been taken into account in the
analysis because of the inevitable increase in its conclusions uncertainty.

\subsection{Fitting conditions}

In first of the tested by us variants, the
$E_n=0.25g(\Delta_0^2-\Delta_n^2)+\delta e_{n}$, functional dependence suggested
by A.V.Ignatjuk and Yu.V.Sokolov (see [24]) has been used.
It is based on the idea of existence of $0.25g\Delta_0^2$ condensation energy
in a nucleus at its transition from normal to superfluid state.
The maximal number of decayed pairs in this variant is limited to value $N=5$,
basically, by possibilities of used fitting algorithm.
The essential increase in $N$ could reduce  $\chi^2_f$ value at least up to
equaling it to $\chi^2_s$ value of the second variant.
But it may be necessary to insignificantly change the obtained conclusions.
For the present the parameter $\delta e_{n}$ for next decayed pair is determined
only [2] in modeling. However, for comparison of (4) with the experiment it was
selected so that to provide the best $\rho$ fitting for the experimental values
[20] (when they are absent - the data on level density from [11] are used).
Correlation function $\Delta_0$ of the ground state of the even-even nuclei is
accepted as equal to the experimental (determined from atomic masses) pairing
energy of last neutron; whereas in odd-neutron nuclei $\Delta_0=12.8/\sqrt A$
approximation was used. Energy dependence of correlation function $\Delta_n$
for the Cooper pair of number $N$ at nucleus excitation energy $U$ was set
within the framework of its theoretically determined and presented in [2]
approximation. It assumes that $n$-quasiparticle ($n=2N$) excitations can exist
above the threshold energy $U_{th}$.
It may be defined by the following expressions:
\begin{eqnarray}
\frac {U_{th}}{C}=3.144 (n/n_c)-1.234(n/n_c)^2~~{\rm for}~ n/n_c \leq 0.424\\
\frac {U_{th}}{C}=1+0.617(n/n_c)^2~~~~~~~~~~~~~~~~{\rm for}~ n/n_c > 0.424
\end{eqnarray}

Here, $C=0.25g\Delta_0^2$ is the condensation energy, and $n_c=0.791g\Delta_0$
is the number of excited quasi-particles in the vicinity of the phase-transition
point from the superfluid to normal state.

The energy dependence of the pairing-gap parameter $\Delta_n$, that is required
for calculations of the level density (4), can be parameterized in the form:
\begin{eqnarray}
\frac {\Delta_n}{\Delta_0}=0.996-2.36(n/n_c)^{1.57}/(U/C)^{0.76}~{\rm for}~ 
U/C\ge 1.03+2.07(n/n_c)^{2.91}\\
\frac {\Delta_n}{\Delta_0}=0  ~~~~~~~~~~~~~~~~~~~~~~~~~~~~~~~~~~~~~~~~~~~~
\rm otherwise
\end{eqnarray}

To correctly account the spin dependence of the level density is a serious
problem because of the ambiguity of [3] model notions of the spin cutoff factor.
In both described here variants, the functional dependence [27] suggested by Fu,
has been used.
The corresponding function, written in language FORTRAN77, is taken from file
RIPL-2 [3] with necessary parameters.
As the breaking threshold for number $N$ pair was selected only by comparing
values of expression (4) with the experimental density of levels, so its value,
suggested in [27], has been replaced with the best one obtained by us.
Values used in calculation of constants $g$ and the best fitting parameters
$E_{N}$ (for $E_N=U_{th}$) of expression (4), found in the first variant, are
presented in tables 1 and 2.

First of all, it is found out that in this representation the experimental level
density requires accounting for not less than 5 partial densities for its
reproduction. According to the notions [25], the effect of pairing takes a
share of nucleus excitation energy, which is equal to $2\Delta_0$.
For five breaking nucleon pairs the total energy of pairing is equal
approximately to 10 MeV for the heaviest of the nuclei that are included in the
analysis. This energy is more than the neutron binding energy and, consequently,
notions [24,28] together with the data such as obtained in [5] seriously
disagree with our experimental data.

The substantial problem is also presented by discrepancy of energy thresholds
of various pairs' breaking. As a rule, for the investigated nuclei this
threshold for the 5-th pair is significantly less than for the 4-th.
Sometimes inversion is observed for other pairs as well.
In some of the nuclei still more inversion of disintegrated pairs' thresholds
is observed. The trivial explanation can consist of a divergence of model
notions [2] and the experiment in close vicinity to the quasiparticles
pair-production thresholds. If it is so, results of the executed approximation
of the experimental level density in the first variant analysis show that with
the error of about 1 MeV of breaking energy of 3, 4 and 5-th nucleon pairs
practically coincide.
Within the framework of notions [2] about the form of Cooper pairs correlation
functions energy dependence of excited nucleus, a picture, corresponding to the 
basic statement [28], is observable.
This statement deals with the generalized model of a superfluid nucleus -- its
phase-transition between superfluid and normal states (but at the essentially
smaller energy of such transitions).

Model approximation of correlation function $\Delta_n$, presented in [2], is
obtained on the base of the experimental data such as [5].
Their basic distinctive feature is that the speed of increase in the level
density is smaller as compared to data of [11,20].
It is true at least for excitation energy some higher than 0.5$B_n$.
In this variant of model notions, at the increase in excitation energy the
smooth enough change in $\Delta_n$ can provide the greater $d\rho/dU$
value only when the expression (4) accounts for five and more breaking pairs.
An alternative opportunity consists in using other ideas of nucleon correlation
functions of Cooper pairs in the excited nucleus.

The smaller values of $n$ can be obtained only with the more rapid decrease
in $\Delta_n$, than it is predicts by eq. (7), as increasing of the
excitation energy $U$. And equation (4) cannot present any other opportunity.
The number of variants of the functional dependences satisfying this condition
is great. In the second variant of the analysis for $E_n$ the following
functional dependence has been used:
\begin{equation}
U-E_n=U-\Delta_0 ln[(U-U_{th})/(p\Delta_0)].
\end{equation}
The only ground for using function (9) for this purpose is a logarithmic
dependence of macrosystem thermal capacity in a point of second-order phase
transition on its temperature. But such dependence is shown only in an ideal
case. In case of a mixture of helium isotopes, for example, the maximal thermal
capacity decreases, whereas the degree of change increases at the increase of
$^3$He concentration.
Therefore, dependence (9) can be accepted as the utmost possible  $\Delta_n$
estimation for pair number $N$ at the energy $U$ with an additional condition
[25] that maximum $E_n$ value is in $E_n \leq n\Delta_0/2$.
Probably, following this condition for the majority of nuclei gives
overestimated $E_n$ values. Only for $^{74}$Ge, $^{185}$W, $^{192}$Ir
and $^{196}$Pt, it is required to increase the maximal value of $E_n$ in (9) by
1.1 - 1.5 times in order to obtain the minimum possible $\chi^2_s$.
Such correction is essential only for the second breaking Cooper pair.
If presence of systematical errors in $\rho$  is taken into account, the
specified $E_n$ increase does not indicate its excessive divergence in various
nuclei. It is also true for the known significant fluctuations of pairing energy
of last nucleons pair. 

The best value of parameter $p$ for all tested nuclei is about 2.2-2.3.
Therefore, in the second of the investigated variants, the pairing influences
the level density $\rho_{qp}(U,J,\pi)$ for next breaking Cooper pair only in
essentially limited, as compared to (7), energy interval
$U_{th} < U \leq U_{th}+p\Delta_0$. I. e., it is practically equal to the known
value of gap width in the low-lying states spectrum of an even-even nucleus.

In the carried out variants of the analysis, the coefficient $K_{coll}$ has
been accepted as independent on excitation energy $U$, spin and parity of levels
and unchanging at the increase of nucleus excitation energy.
Its absolute value is almost completely determined by a ratio of the
experimental level density and the density of two or three quasiparticle
excitations. In a case, when this value for any number $N>1$ Cooper pairs differs
by some times, such discrepancy is easily compensated by changes in $U_{th}$
within the limits of, several hundreds keV maximally.

At the analysis of the experimental data in even-odd compound nucleus some
ambiguity arises at a choice of threshold of 3-quasiparticle level excitation
energy. If the threshold of their production is accepted as sufficiently high
(about several MeV), in such nuclei, the required collective enhancement factor
of one-quasiparticle level density will exceed similar value for even-even and
odd-odd nuclei by a factor of some tens. Therefore, it has been postulated,
that in the excitation energy interval of about 2-3 MeV above the ground state
of even-odd nucleus, the base density is the one of three-quasiparticle
excitations instead of the density of one-quasiparticle ones. 

In odd-odd compound nuclei the density of 2-quasiparticle excitations assumes
that there is an excitation both of neutron, and proton quasiparticles.
The type of quasiparticles for 4 and more quasiparticle excitations in these
nuclei, as well as in even-even ones cannot be established by the analysis
carried out here.

\subsection{Results and their discussion}

Examples of the best approximation of the experimental data [11,20] for 4
nuclei with various parity of neutrons and protons are displayed in Fig. 2.
It also shows the partial level density obtained in the first variant of the
analysis. In Figs. 3-8 the similar data are presented for the most of the
analyzed nuclei with partial densities obtained in the second variant of
analysis. In tab. 2 the parameters of the approximated partial level density
for the second variant of accounting for nucleon pairing interactions in
nucleus are presented.
Thresholds of excitation energy $U_{th}$, necessary for calculation by
expressions (4) and (9), are averaged in it after division of $E_N$ onto the
approximated value of correlation function $\Delta_0=12.8/\sqrt A$
for nuclei with different neutron and proton parity, separately.

     From these data quite unequivocally follows, that:
     
1. The first step-like structure in level density [11,20] is caused by
existence of, at least, two quasiparticles in nucleus of any type;

2. For the precise reproduction of the level density (comparable with an
experimental data error) it is required to postulate the breaking from three
up to five and more such pairs; 

3. 2-3 and more nucleon Cooper pairs can have practically equal breaking
threshold if only the effect of nucleons pairing of any pair demonstrates
itself in a wide [2] interval of nucleus excitation energy.
If not to take into account the circumstance that the approximation of
experimental data on $\rho$  both from [11] and [20] necessarily inverts
thresholds of breaking at least of the fourth and fifth pairs, the proximity
of their values allows one to speak about practically observable simultaneous
breaking of several Cooper pairs.
That is, about [28] a nucleus phase-transition from superfluid to the normal
state;

4. The coefficient $K_{coll}$, taken from the data [20], has practically equal
value in both even-even and even-odd compound nuclei and is significantly higher
in odd-odd ones (for the level density from a variant of the analysis [11]).
This difference is qualitatively explained by value and error distinction sign
of the level density given by [11] as regard to [20];

5. From the data of approximation, presented on Figs. 3-8, it is visible,
that in the second variant of the analysis the experimental level density can
be well reproduced at the account of only 6 or 7-quasiparticle excitations.
Moreover, representation of correlation functions as (9) allows one to obtain
small enough value of the level density of $n$-quasiparticle excitations,
having lower energy than $U_{th}$.
So, the known from nuclear spectroscopy fact of the 2-quasiparticles excited
levels presence within the limits of nucleus excitation energy somewhat lower
than $2 \Delta_0$ is thus explained;

6. In Tables 1 and 2 data on comparison of average values of collective
enhancement factor of the level density are collected and show smaller scatter
in the second variant of the analysis. 
In this variant differences $<E_2/\Delta_0> - <E_1/\Delta_0> \approx 4$
and $<E_3/\Delta_0> - <E_2/\Delta_0> \approx 2$ coincide in nuclei of various
types within the limits of an error;

7. In [25] it is predicted that the level density of odd A nucleus corresponds
to that of even nucleus with excitation energy $U+\Delta_0$, accordingly.
Results of the first variant of the analysis (Tab. 1) do not prove this
prediction to be true. However, in the second one it is carried out within the
limits of $E_n$ energies determination error; 

8. The shell effects demonstrate itself to the maximum extent in near magic
nuclei being close to $N$=82 and 126.
Their exhaustive reproduction is impossible within the framework of
expression (4). More exact approximation can demand to revise the basic
assumption [25] about equidistant character of single-particle spectrum of
near magic nucleus and to account [24] for shell inhomogeneities of
single-particle spectrum.
It may also be required to account for change in $K_{coll}$ value for different
nuclei and excitation energies, not taken into account by performed analysis.

Therefore, it is undoubtedly required to obtain better experimental data on
gamma-gamma coincidences and further development of algorithms of their
analysis. First of all it concerns the determination of $E_n$ energy values,
connected with collective type excitations near a breaking threshold of pair
number $N$. The problem of $K_{coll}$ value determination at a given excitation
energy $U$ is directly connected to this circumstance.
For even-even nuclei it is visible from the given tables and figures.
Approximation of $\rho$  is done in them starting with $U \approx 2$ MeV.
Just this is the explanation of the discrepancy of $E_1$ energy value with
$2\Delta_0$. As long as in even-even nuclei of $\Delta_0$ region there are only
levels of vibrational type, so far it is necessary to accept
$K_{coll} \approx \infty $ in this excitation region.
The available experimental data, unfortunately, do not give an opportunity to
estimate this problem near to values $E_2$ and $E_3$ basically because of the
lack of theoretical representations on this account and mathematical methods
of the analysis of the experimental data.

Despite of the specified ambiguity, volume and quality of the obtained [11,20]
experimental data presents extremely favorable possibilities for the greatest
possible development of modern models of the level density.
The obtained values of the most probable parameters of the experimental data
approximations can be used for $\rho$ calculation in nuclei that are relatively
far from the magic ones. It may be done either directly or at interpolation of
the data given by Tables 1 and 2, or by using the corresponding average values.
It is certainly necessary to develop new model ideas of the nucleon correlation
functions energy dependence for near magic nuclei in the whole region of
$E_{ex} <B_n$ and about $\rho$ vibration enhancement factor. 

It is also necessary to reveal the degree with which the initial levels
$\lambda$
structure influences on the two-step cascade intensity and to estimate the
$\rho$ and $k$ systematical error connected to this circumstance. 
     
\section{Conclusion}

There is an urgent need for both the experimental and the theoretical
determination of the excited level density in a given nuclear reaction, as well
as the probability of its products emission for all possible excitation energies
with a guaranteed systematical error not exceeding several tens of percents.
That is, it is to be accomplished with the same accuracy of measuring various
spectra of nuclear reactions and their cross sections, as carried out,
presently. Only such reduction of available systematical errors of the
experimental $\rho$ and $k$ determination allows one to observe dynamics of
nucleus structure change at its excitation energy variation and to reveal
factors causing it. There is no other possibility to develop new nucleus model
and to specify the existing ones. Known techniques [5,7,8] of $\rho$ and $k$
determination are incapable of solving this problem.
It is true either because of the impossibility of getting accuracy-specified
model-designed transmission coefficient for nuclear reaction's nucleon products,
or because of practical inaccessibility of the required measurement accuracy of
gamma-rays spectra, accompanying nuclear reactions with the charged particles
emission.

Specific feature of the model-free simultaneous extraction [8,11,20] of both
the level density and strength functions of emission of the nuclear reaction's
registered products is necessary to determine the intervals of their values,
reproducing experimental spectra with the required accuracy.
Essentially, it is provided for by the irremovable ambiguity of such problem's
solution both for [8] and [11,20] techniques.
The marked circumstance is connected with degeneracy of the equation systems,
connecting values of the sought parameters with the measured spectra
intensities.
Model notions and hypotheses (first of all, on independence of strength
functions on a nucleus's excited levels structure) reduce or fully eliminate
[5] the equation systems degeneracy.
But they necessarily lead to an unknown systematical error $\rho$ and $k$.
Therefore, reliable results for $\rho$ and $k$ can be only obtained by using
a mathematically correct technique of the experimental data processing.
The data should most precisely reflect this circumstance.

The accumulated, up to the present, data file on two-step cascades following
thermal neutron radiative capture, points to the existence of systematical
discrepancy in experimentally obtained data for $\rho$ and $k$ and their
model-calculated values. These data also specify excessive idealization of
notions about a nucleus, used by techniques of [5,7,8] type for the extraction
of its major parameters. Presence of significant systematical discrepancy
between the experiment and theory allows one to hope on the possibility of a
substantial improvement in model notions about properties of a nucleus at its
excitation being lower than nucleon binding energy.

\newpage
\begin{center}{References} \begin{flushleft} \begin{tabular}{r@{ }p{5.65in}} 
$[1]$ & V.M. Maslov, Nucl. Phys. A. 743 (2004)  236.\\
& V.M.  Maslov, Yu.V.  Porodzinskij, M.  Baba, A.  Hasegawa, Nucl.  Sci. 
Eng. 143 (2003) 1.\\ 
$[2]$ & F.  Rejmund, A.V. Ignatyuk, A.R. Junghans, K.-H. Schmidt, Nucl. Phys. 
A678 (2000) 215.\\ 
$ [3]$ & Reference Input Parameter Library RIPL-2.  Handbook for
 calculations of nuclear reaction data.  IAEA-TECDOC, 2002, http://www-
nds.iaea.or.at/ripl2/ \\
 & Handbook for Calculation of Nuclear Reactions Data, IAEA, Vienna,
 TECDOC-1034, 1998.\\ 
 $ [4]$ &  V.A. Khitrov, Yu.P. Popov, A.M. Sukhovoj, Yu.S. Yazvitsky,
 in Proceedings of III International symposium
on neutron capture gamma-ray spectroscopy,
ed. by R.E. Chrien,   W.R. Kane, NY and London, Plenum Press, 1979, 665.\\
 $[5]$ & O.T.  Grudzevich et. al., Sov.  J.  Nucl.  Phys. 53 (1991) 92.\\
 & B.V.  Zhuravlev, Bull. Rus. Acad. Sci. Phys. 63 (1999) 123.\\ 
 $[6]$ & V. Bondarenko, J. Honzatko, I. Tomandl et al.,
Nucl. Phys. A. 762 (2005) 167.\\
 & J. Honz\'atko et al., Nucl.  Phys.  645 (1999) 331.\\
 $[7]$ & G.A.  Bartholomew et al., Advances in nuclear physics 7 (1973) 229.\\ 
 $[8]$  & A.  Schiller et al., Nucl.  Instrum. Methods Phys.  Res., A447 (2000) 
498.\\
 $[9]$ & A.M.  Sukhovoj, V.A.  Khitrov, Li Chol, in Proceedings of
 XII International Seminar on   Interaction of Neutrons with Nuclei,
E3-2004-169, Dubna, 2004, p.438, nucl-ex/0409016\\
 $[10]$ & V.  G.  Soloviev et al., J.  Phys.  G., Nucl.  Phys.  20 (1994) 113.\\ 
 & V.  G.  Soloviev et al., Part.  Nucl.  27(6) (1997) 1643.\\ 
 & V.  G.  Soloviev,
 Theory of atomic Nuclei:  Quasiparticles and Phonons, Institute of Physics 
Publishing,
 Bristol and Philadelphia, 1992.\\ 
 & V.  G.  Soloviev, Theory of Complex Nuclei, Pergamon Press, Oxford, 1976.\\
 $[11]$ & E.V.  Vasilieva, A.M.  Sukhovoj, V.A.  Khitrov, Phys.  At. Nucl. 
 64(2) (2001) 153, nucl-ex/0110017\\
 $[12] $ & S.T. Boneva, V.A.  Khitrov, A.M.  Sukhovoj, Nucl.  Phys.  A589 
(1995) 293.\\
  $[13]$ & A. Voinov et al., Nucl.\ Instrum.\ Methods Phys.\ Res.\ A497 (2003) 
350.\\
 & A.  Voinov  et al.\rm, Yad.  Fiz. 67(10) (2004) 1891.\\
 & F.  Be\v{c}v\'{a}\v{r} et al., Phys.\ Rev.\ C  52\rm, (1995) 1278.\\
\end{tabular}\end{flushleft} \end{center} \newpage \begin{center}
\begin{flushleft} \begin{tabular}{r@{ }p{5.65in}}
 $[14]$ & Yu.P.  Popov, A.M.  Sukhovoj, V.A.  Khitrov, Yu.S.  Yazvitsky, Izv. AN 
SSSR,
 Ser.  Fiz.  48 (1984) 1830.\\
 $[15]$ &A.M.  Sukhovoj, V.A.  Khitrov, Physics of Atomic Nuclei 62(1) (1999) 
19.\\
 $[16]$ & M.A. Lone, R.A. Leavitt and D.A. Harrison, Nucl.  Data 
Tables 26(6) (1981) 511. \\
 $[17]$ & http://www-nds.iaea.org/pgaa/egaf.html\\
 $[18]$ & V.A Khitrov, Li Chol, A.M. Sukhovoj,
 in Proceedings of  XI International Seminar on Interaction of Neutrons with 
Nuclei, Dubna, 22-25
 May 2003, E3-2004-9, Dubna, (2004) 98, nucl-ex/0404028\\\
 $[19]$ & V.A.  Bondarenko et all, in Proceedings of XII
 International Seminar on Interaction of Neutrons with Nuclei,
 E3-2004-169, Dubna, 2004, p.  38, nucl-ex/0406030, nucl-ex/0511017\\
 $[20]$ &  A.M.  Sukhovoj, V.A.  Khitrov, Physics of Elementary Particles
and Atomic Nuclei  36(4) (2005) 697. (in Russian)\\
 $[21]$ & http://www.nndc.bnl.gov/nndc/ensdf.\\
 $[22]$ & S.F.  Mughabghab, Neutron Cross Sections
 BNL-325.  V.  1.  Parts A, B, edited by Mughabhab S.  F., Divideenam M., Holden 
N.E., N.Y. Academic Press, (1984)\\
 $[23]$ &  A.M.  Sukhovoj, V.A.  Khitrov, E.P.  Grigoriev, Report INDC(CCP),
  Vienna, 432 (2002) 115, nucl-ex/0508007, nucl-ex/00508008.\\
 $[24]$ & A.V.  Ignatyuk, Report INDC-233(L) (IAEA Vienna 1985)\\
 $[25]$ & V.M.  Strutinsky,  in Proc. of  Int.  Conf.  Nucl. Phys., Paris (1958) 
617.\\
 $ [26]$  & E.  B\v{e}t\'ak, P.  Hodgson, Rep.  Progr.  Phys. 61 (1998) 483.\\
 $[27]$ & C.Y. Fu, Nucl.  Sci.  Eng. 92 (1986) 440.\\
 $ [28] $ & E.M. Rastopchin, M.I. Svirin, G.N.  Smirenkin, Yad. Fiz. 52 
(1990)1258.\\
 $[29]$ & W. Dilg, W.  Schantl, H.  Vonach, M.  Uhl, Nucl.  Phys. A217 (1973) 
269.\\
\end{tabular}\end{flushleft} \end{center}
\newpage
{\sl Table 1.\\ }  The best values of the collective enhancement factor $K_{coll}$ for
analyzed nuclei, $E_N$ is the value of breaking thresholds of pair number $N$,
MeV.
 \begin{tabular}{|r|r|r|r|r|r|r|r|} \hline 
 Nucleus& $g$, MeV$^{-1}$ &$K_{coll}$& $E_1$& $E_2$& $E_3$ & $E_4$& $E_5$\\\hline 
 $^{40}$K    &4.04  &15.0 & 0.20 & 5.43 & 5.42 & 3.33 & 2.13\\
 $^{60}$Co   &4.04  &11.6 &-1.88 & 0.14 & 4.38 & 4.32 & 2.09\\
 $^{80}$Br   &6.18  &14.0 &-2.10 & 4.05 & 4.83 & 2.56 & 1.46\\
 $^{128}$I   &7.79  &33.0 &-1.75 & 2.15 & 1.95 & 3.05 & 2.45\\
 $^{140}$La  &7.79  &17   &-1.10 & 2.37 & 1.98 & 2.29 & 1.79\\
 $^{160}$Tb  &10.7  &22.6 &-1.11 & 2.12 & 1.78 & 2.46 & 1.86\\
 $^{166}$Ho  &10.3  &21   &-1.71 & 1.95 & 1.83 & 2.62 & 1.80\\
 $^{170}$Tm  &10.7  &27.3 &-1.69 & 2.04 & 2.19 & 2.65 & 2.05\\
 $^{176}$Lu  &11.0  &23.8 &-1.62 & 1.95 & 1.90 & 2.63 & 2.04\\
 $^{182}$Ta  &10.9  &32.8 &-1.40 & 1.84 & 1.67 & 2.87 & 2.27\\
 $^{192}$Ir  &12.3  &27.8 &-0.97 & 1.79 & 2.07 & 2.80 & 1.90\\
 $^{198}$Au  &9.88  &6.9  &-1.64 & 1.82 & 2.03 & 2.37 & 1.48\\
 \hline
  &&21(8)&&&&&\\\hline 
 $^{74}$Ge  &6.05  &5.5  &-0.14 & 3.59 & 4.27 & 4.24 & 5.04\\
 $^{114}$Cd &7.90  &17.8 &-0.51 & 2.49 & 3.14 & 3.52 & 4.10\\
$^{118}$Sn  &8.08  &1.5  &-0.58 & 3.17 & 2.96 & 2.75 & 2.41\\
$^{124}$Te  &8.27  &8.8  &-0.13 & 2.42 & 3.16 & 3.91 & 4.31\\
$^{138}$Ba  &6.98  &1.5  &-0.72 & 2.48 & 3.27 & 3.56 & 2.56\\
$^{150}$Sm  &10.3  &10.0 &-1.12 & 2.04 & 2.79 & 3.29 & 2.80\\
$^{156}$Gd  &10.2  &19.6 &-0.87 & 1.66 & 3.02 & 3.60 & 3.21\\
$^{158}$Gd  &10.1  &13.  &-0.82 & 1.96 & 3.06 & 3.32 & 2.92\\
$^{164}$Dy  &9.26  &9.4  &-1.43 & 1.76 & 2.51 & 2.81 & 2.31\\
$^{168}$Er  &10.1  &14.  &-0.82 & 1.67 & 2.48 & 3.45 & 3.04\\
\hline
\end{tabular}
\newpage
\begin{tabular}{|r|r|r|r|r|r|r|r|} \hline 
$^{174}$Yb  &9.70  &9.5  &-1.32 & 1.74 & 2.56 & 2.78 & 2.28\\
$^{184}$W   &11.0  &7.0  &-0.69 & 1.97 & 2.15 & 3.02 & 2.51\\
$^{188}$Os  &11.7  &13.  &-0.91 & 2.18 & 2.31 & 3.59 & 3.19\\
$^{190}$Os  &11.4  &5.9  &-1.04 & 2.44 & 3.12 & 3.18 & 2.54\\
$^{196}$Pt  &9.30  &25.  &-0.85 & 2.04 & 3.11 & 3.22 & 3.26\\
$^{200}$Hg  &7.30  &3.3  &-1.34 & 3.13 & 2.69 & 2.16 & 3.46\\
\hline
& &10(6)  &&&&&\\\hline 
$^{177}$Lu  &11.2  &11.7  & -1.04 & 1.59 & 2.58 & 1.46&\\\hline   
$^{71}$Ge   &5.06  & 6.2 &-1.49 & 1.17 & 0.18 &-0.49&\\
$^{125}$Te  &8.45  &12.4 &-1.14 & 1.38 & 1.08 & 0.78&\\
$^{137}$Ba  &8.42  & 6.5 & 0.69 & 2.44 & 2.47 & 1.17&\\
$^{139}$Ba  &8.70  & 0.9 &-1.89 & 1.98 & 1.12 &-0.28&\\
$^{163}$Dy  &9.47  &11.  &-1.23 & 2.04 & 0.84 & 0.11&\\
$^{165}$Dy  &9.37  & 2.  &-1.59 & 1.58 & 0.00 &-0.80&\\
$^{181}$Hf  &10.7  &9.9  &-1.48 & 1.35 & 0.96 & 0.36&\\
$^{183}$W   &10.2  &10.9 &-1.62 & 1.95 & 1.50 & 0.32&\\
$^{185}$W   &10.3  &4.1  &-1.44 & 1.79 & 0.74 &-0.49&\\
$^{187}$W   &11.3  &15.5 & 0.17 & 1.63 & 0.94 & 0.86&\\
$^{191}$Os  &10.4  &13.0 &-1.62 & 1.09 & 0.74 & 0.14&\\
$^{193}$Os  &10.2  &11.8 &-1.23 & 1.61 & 0.84 & 0.17&\\
\hline
 &&8.4(45)&&&&&\\ \hline 
\end{tabular} 

\newpage
{\sl Table 2.\\ }  $\Delta J$ is the interval of spins for which the
experimental value $\rho$ is determined, parameter $p$ of expressions (9)
and the best values of the collective enhancement factor $K_{coll}$ for
analyzed nuclei in the second variant of the analysis.
$E_N$, MeV is the value of breaking thresholds of pair number $N$.
$R =\chi^2_s/\chi^2_f$ - relations of criteria of fitting quality of the second
(s) and the first (f) variants of the analysis.

\begin{tabular}{|r|r|r|r|r|r|r|r|} \hline 
   Nucleus&$\Delta J$ &p &$K_{coll}$& $E_1$& $E_2$& $E_3$&  $R$ \\\hline 
 $^{40}$K   &1-3 &2.2   &3.9  & -1.45 & 2.5  & 7.1 & 0.6\\
 $^{60}$Co  &2-5 &2.2   &5.6  & -2.10 & 5.17 & 6.2 & 0.4\\
 $^{80}$Br  &0-3 &2.2   &8.8  & -2.50 & 3.45 & 6.0 & 0.5\\
 $^{128}$I  &1-4 &2.2   &26.  & -1.24 & 1.95 & 4.8 & 3.3\\
 $^{140}$La &3-5 &2.2   &11.7 & -1.24 & 1.95 & 4.8 & 0.5\\
 $^{160}$Tb &0-3 &2.0   &10.6 & -1.40 & 2.73 & 4.3 & 0.6\\
 $^{166}$Ho &2-5 &2.2   &11.0 & -1.50 & 2.90 & 4.5  & 0.5\\
 $^{170}$Tm &0-2 &2.2   &16.4 & -1.46 & 2.70 & 5.3 & 0.2\\
 $^{176}$Lu &2-5 &2.3   &13.8 & -1.51 & 2.68 & 4.9 & 0.4\\
 $^{182}$Ta &2-5 &2.2   &18.5 & -1.00 & 2.00 & 5.2  & 0.9\\
 $^{192}$Ir &0-3 &2.2   &16.0 & -1.80 & 3.15 & 4.6  & 0.6\\
 $^{198}$Au &1-3 &2.2   &12.0 & -.96  & 3.81 & 6.1  & 0.5\\
 \hline
  & &2.19  &13(6)& -1.3(3)& 2.6(8) & 4.7(8) & \\\hline 
 $^{74}$Ge  &3-6 &2.2   &5.1  &  1.80 & 6.00 & 9.5  & 0.5\\
 $^{114}$Cd &0-2 &2.2   &7.6  &  0.09 & 3.72 & 7.2  & 0.8\\
$^{118}$Sn  &0-2 &2.2   &2.0  & -0.20 & 4.10 & 4.7  & 0.4\\
$^{124}$Te  &0-2 &2.4   &10.5 & -0.20 & 2.90 & 6.9  & 0.5\\
$^{138}$Ba  &1-3 &2.2   &1.6  &  0.10 & 4.60 & 7.5  & 0.9\\
$^{150}$Sm  &3-5 &2.4   &10.4 & -1.03 & 2.63 & 4.4  & 0.7\\
$^{156}$Gd  &1-3 &2.3   &17.0 & -1.33 & 2.42 & 4.8  & 0.6\\
$^{158}$Gd  &1-3 &2.0   &6.0  & -0.30 & 3.10 & 4.8  & 0.6\\
$^{164}$Dy  &1-4 &2.3   &14.  &  0.15 & 3.33 & 6.5  & 0.9\\
$^{168}$Er  &2-5 &2.3   &8.5  & -0.24 & 3.02 & 6.2  & 1.6\\
\hline
\end{tabular}
\newpage
\begin{tabular}{|r|r|r|r|r|r|r|r|} \hline
$^{174}$Yb  &1-4 &2.2   &12.  & -0.43 & 3.23 & 6.9  & 0.2\\
$^{184}$W   &0-2 &2.3   &3.4  & -0.07 & 3.25 & 4.8  & 0.4\\
$^{188}$Os  &0-2 &2.3   &10.  & -0.26 & 3.76 & 5.8  & 0.7\\
$^{190}$Os  &0-3 &2.2   &10.3 & -0.26 & 3.76 & 5.8  & 1.4\\
$^{196}$Pt  &0-2 &2.3   &36.  & -0.11 & 3.67 & 7.0  & 0.6\\
$^{200}$Hg  &0-1 &2.2   &3.6  &  0.16 & 4.44 & 6.8  & 1.0\\
\hline
 & &2.25 & 10(8)     &0.1(9) &3.5(7)& 5.7(17) &\\\hline 
$^{177}$Lu &11/2-15/2 &2.2 &7.1    & -0.90 & 2.48 & 4.4 & 1.0\\
\hline 
$^{71}$Ge  &1/2-3/2 &2.3   & 9.1 & -2.46 & 2.07 & 5.2 & 1.2\\
$^{125}$Te &1/2-3/2 &2.3   &14.5 & -1.52 & 2.56 & 5.6 & 1.1\\
$^{137}$Ba &1/2-3/2 &2.2   & 5.0 & -2.71 & 3.83 & 4.0 & 0.1\\
$^{139}$Ba &1/2-3/2 &2.3   & 2.7 & -2.86 & 1.63 & 1.8 & 0.6\\
$^{163}$Dy &1/2-3/2 &2.3   &13.  & -0.82 & 2.08 & 3.9 & 2.0\\
$^{165}$Dy &1/2-3/2 &2.3   & 5.6 & -1.06 & 2.49 & 3.6 & 0.5\\
$^{181}$Hf &1/2-3/2 &2.3   &14.8 & -1.60 & 1.70 & 3.8 & 1.2\\
$^{183}$W  &1/2-3/2 &2.2   &16.3 & -1.49 & 2.58 & 4.6 & 0.2\\
$^{185}$W  &1/2-3/2 &2.2   &12.2 & -1.87 & 3.19 & 4.0 & 0.2\\
$^{187}$W  &1/2-3/2 &2.2   & 8.4 & -1.27 & 2.81 & 3.6 & 0.3\\
$^{191}$Os &1/2-3/2 &2.3   &18.7 & -1.30 & 1.53 & 4.0 & 2.0\\
$^{193}$Os &1/2-3/2 &2.3   &14.0 & -0.94 & 2.31 & 3.9  & 0.3\\
\hline  & &2.27   &11(5)& -1.6(5) &2.4(7)& 3.9(8)&\\ \hline
\end{tabular} 

\newpage
\begin{figure}
\vspace{3cm}
\leavevmode
\epsfxsize=14cm
\epsfbox{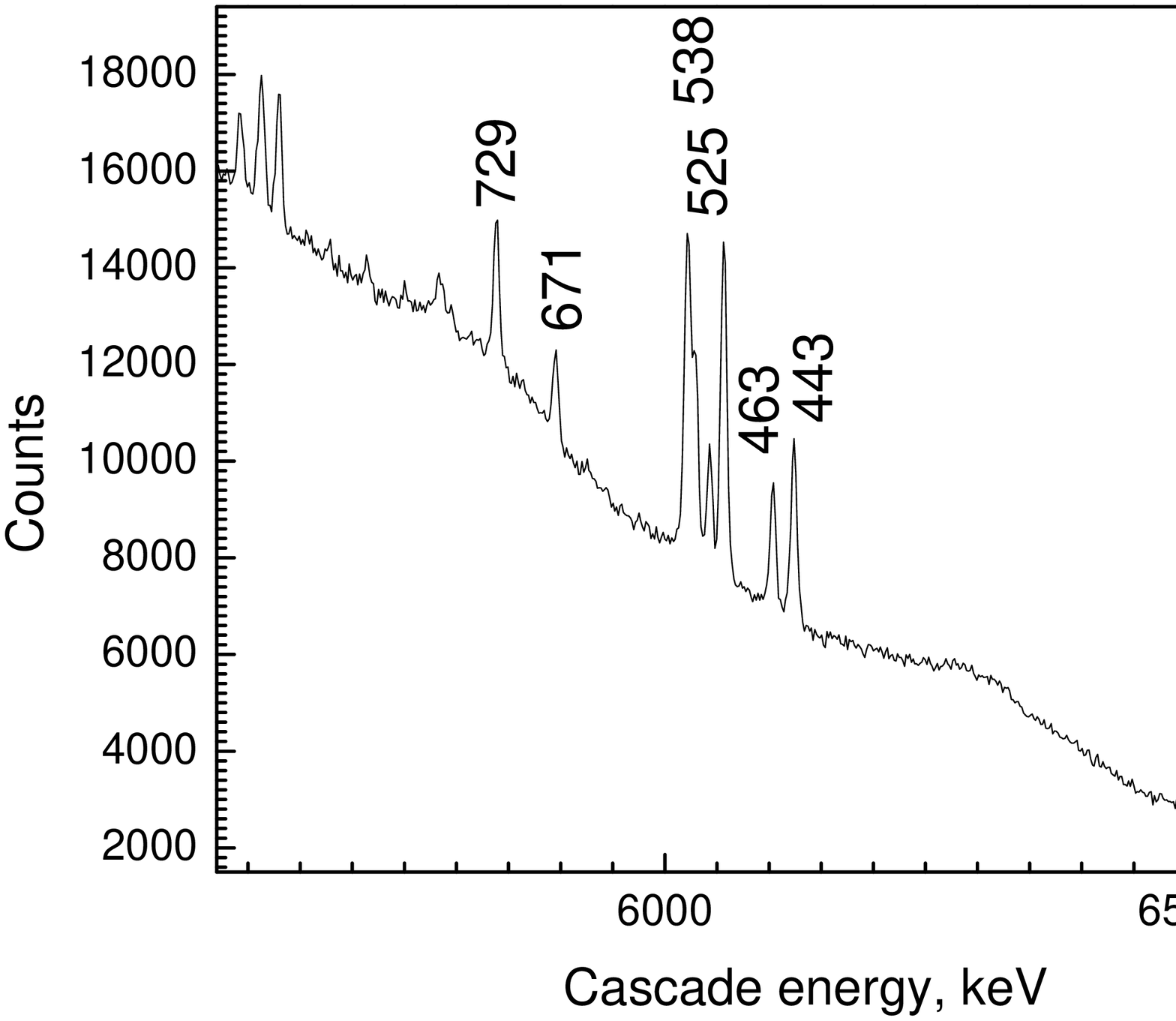}
\vspace{-4.cm}

{\sl Fig.~1.
~It presents the main part of the sum coincidence spectrum for the target
enriched in $^{124}$Te. Full energy peaks are labeled with the energy (in keV)
of final cascade levels.}
\end{figure}

\begin{figure}
\vspace{2cm}
\leavevmode
\epsfxsize=14cm
\epsfbox{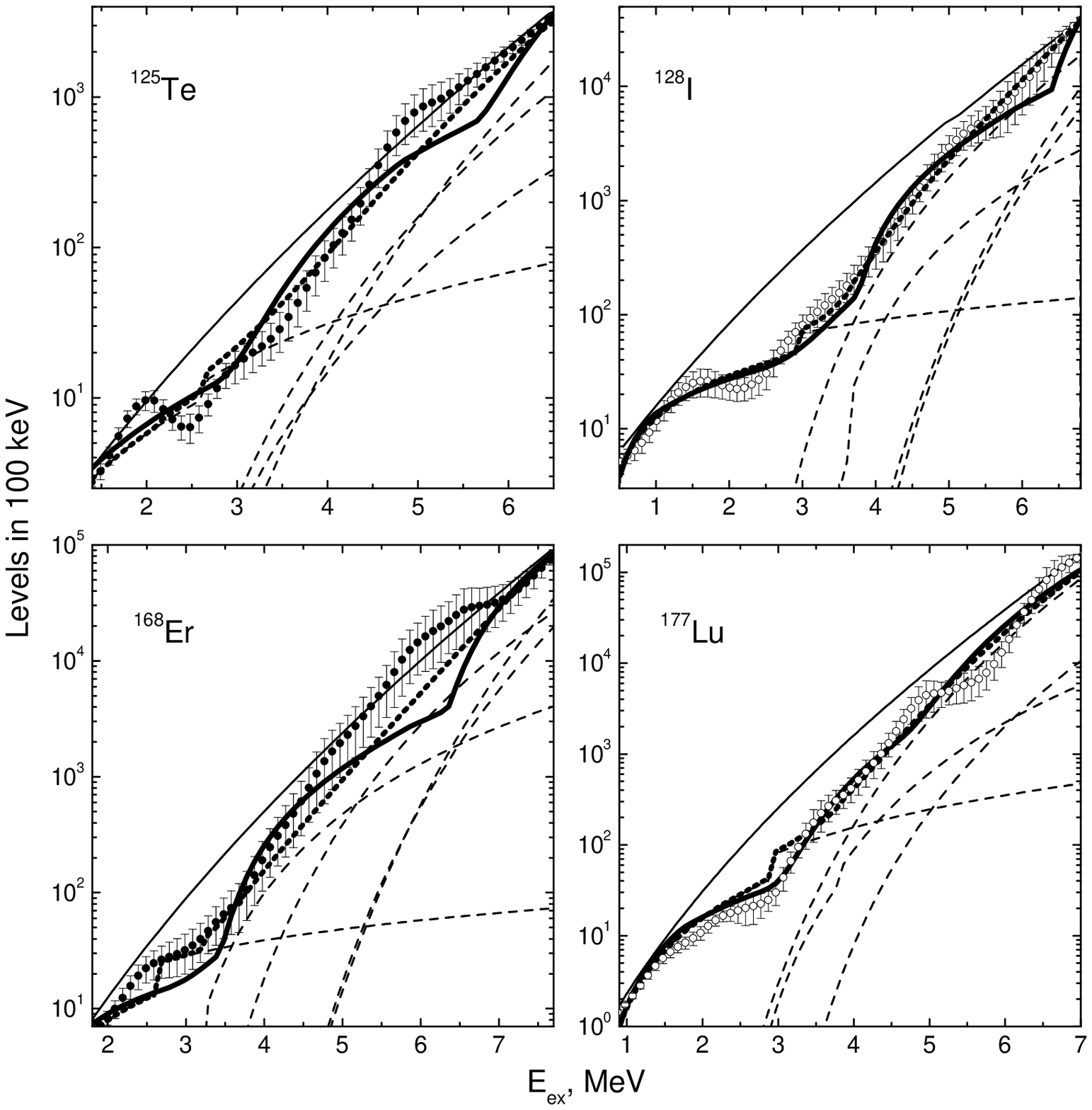}
\vspace{-6.cm}

{\sl Fig. 2.~It shows examples of the experimental data on approximation for nuclei
$^{125}$Te, $^{128}$I, $^{168}$Er and $^{177}$Lu by the partial level density
in the first variant of the analysis.
Full points with error bars represent the experimental data [20], open points
show the data [11]. A thin dotted line indicates partial density, points
display their sum. A solid line presents the sum of the partial level density
from the second variant of the analysis. Thin line stands for the level density 
calculated within model [29].} \\
 \end{figure}
\newpage
\begin{figure}
\leavevmode
\epsfxsize=14cm

\epsfbox{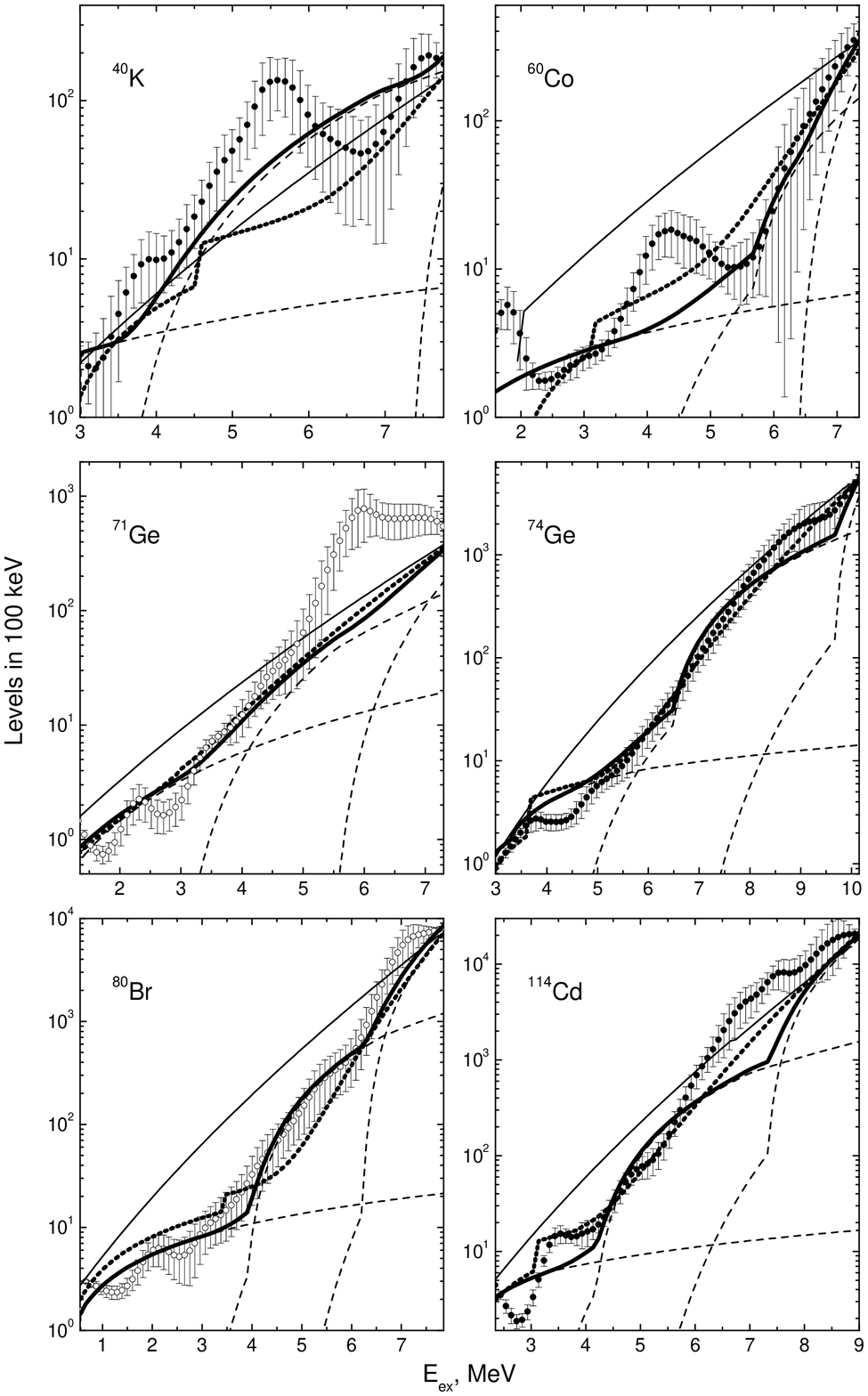}

{{\sl Fig. 3.}
 The same as in Fig. 2 for nuclei: $^{40}$K, $^{60}$Co, $^{71,74}$Ge,
$^{80}$Br, $^{114}$Cd.
Here, the partial level densities are shown as obtained from the second
variant of the analysis.
}
\\\\
\end{figure}
\begin{figure}
\leavevmode
\epsfxsize=14cm

\epsfbox{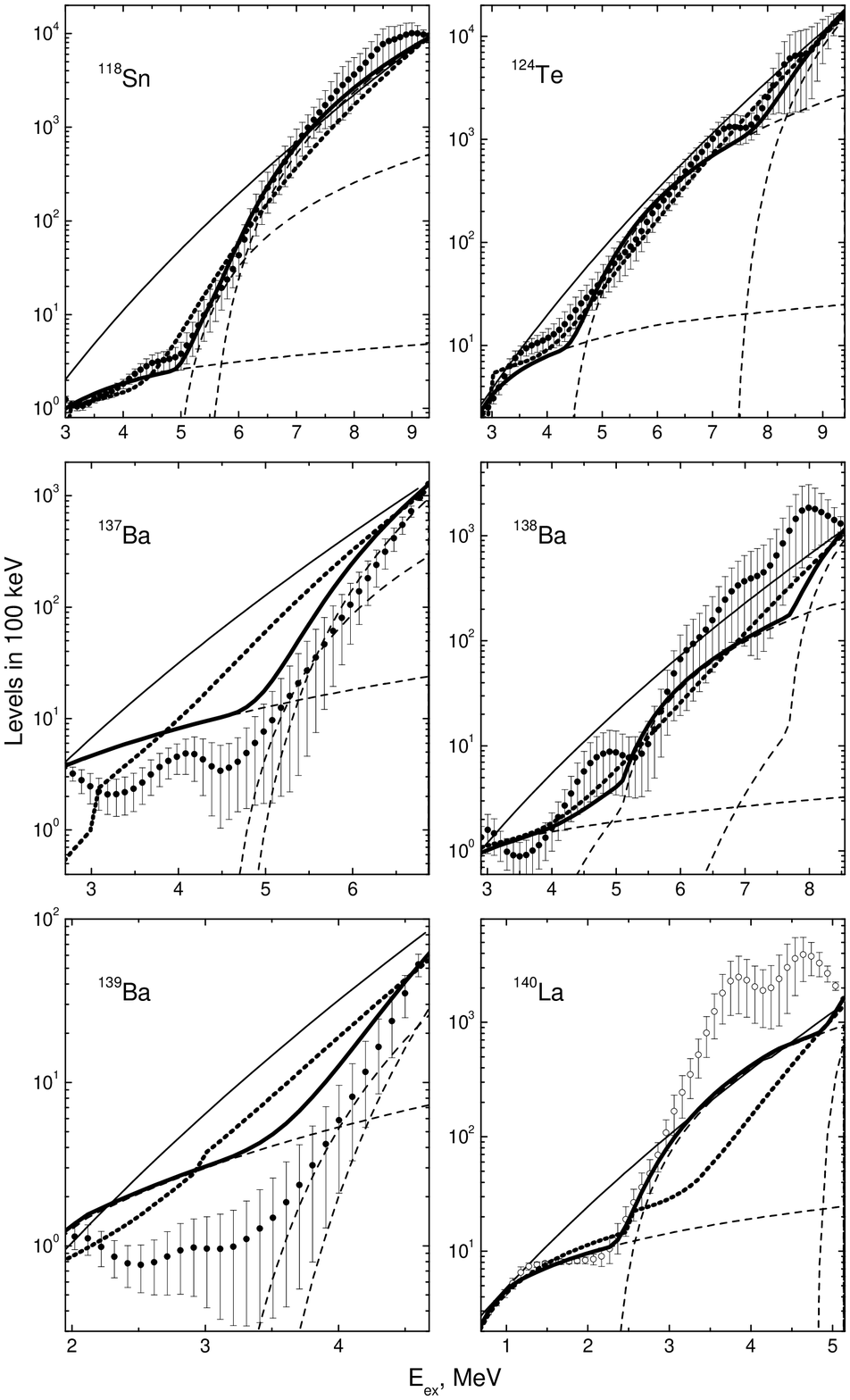}

{{\sl Fig. 4.}
 The same as in fig. 2 for nuclei:  $^{118}$Sn, $^{124}$Te,
$^{137,138,139}$Ba, $^{140}$La. }\\\\\\
\end{figure}
\newpage
\begin{figure}
\vspace{2cm}
\leavevmode
\epsfxsize=14cm

\epsfbox{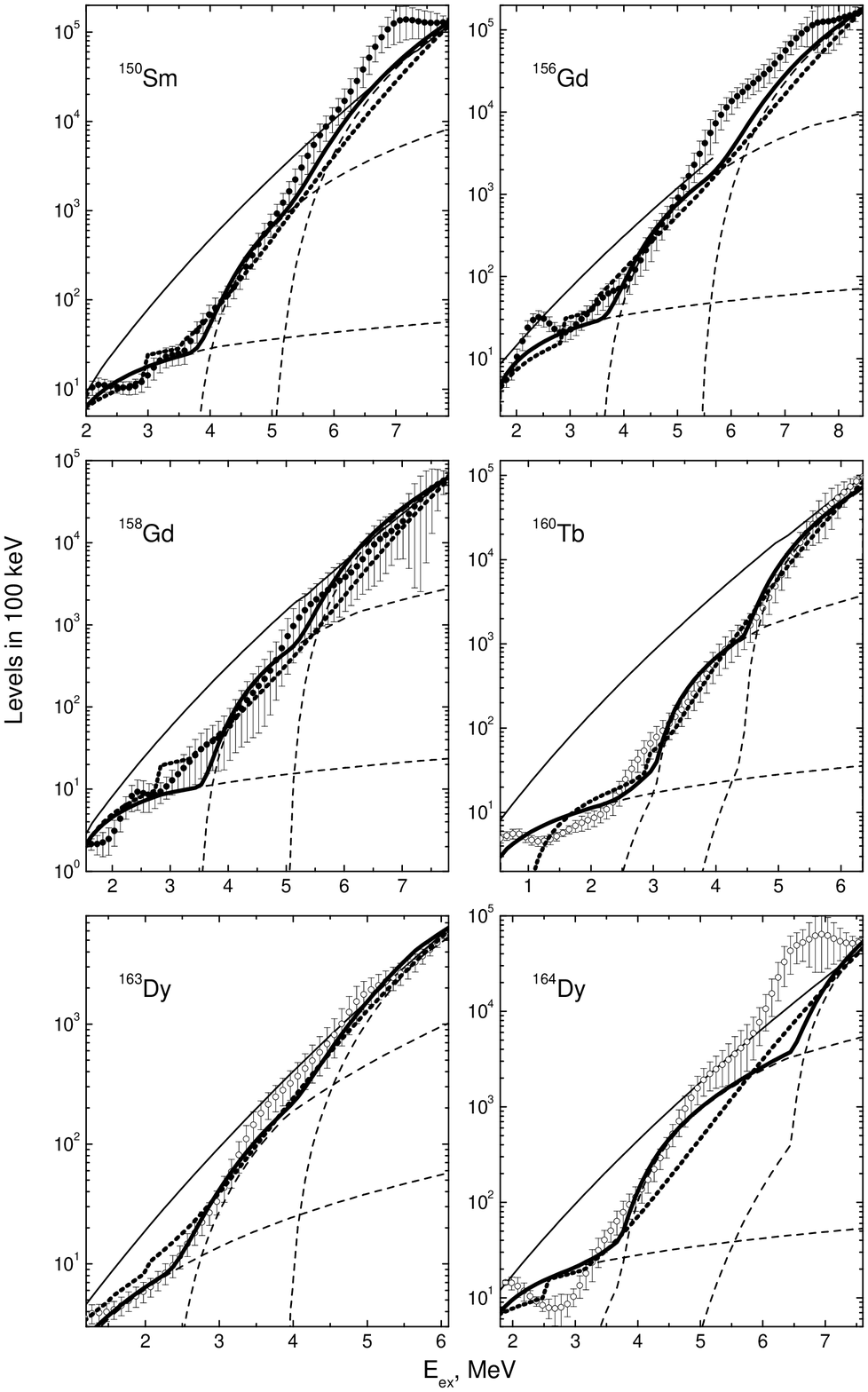}

{\sl{Fig. 5.} The same as in Fig. 2 for nuclei: $^{150}$Sm, $^{156,158}$Gd,
$^{160}$Tb, $^{163,164}$Dy.}\\
\end{figure}
\newpage
\begin{figure}
\vspace{2cm}
\leavevmode
\epsfxsize=14cm

\epsfbox{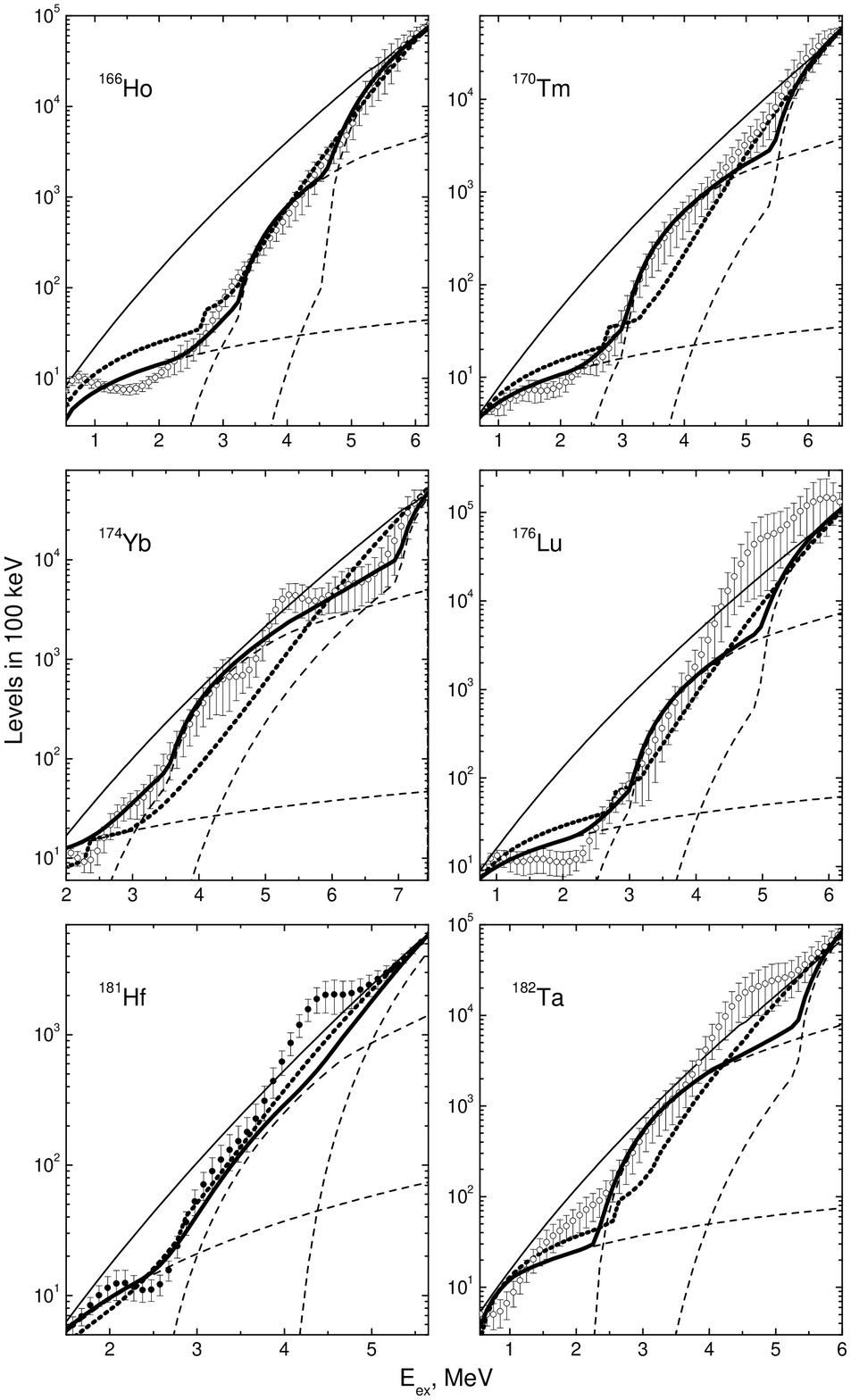}

{{\sl Fig. 6.}
The same as in Fig. 2 for nuclei: $^{166}$Ho, $^{170}$Tm,
$^{174}$Yb, $^{176}$Lu, $^{181}$Hf, $^{182}$Ta.}\\
\end{figure}
\newpage
\begin{figure}
\vspace{2cm}
\leavevmode
\epsfxsize=14cm

\epsfbox{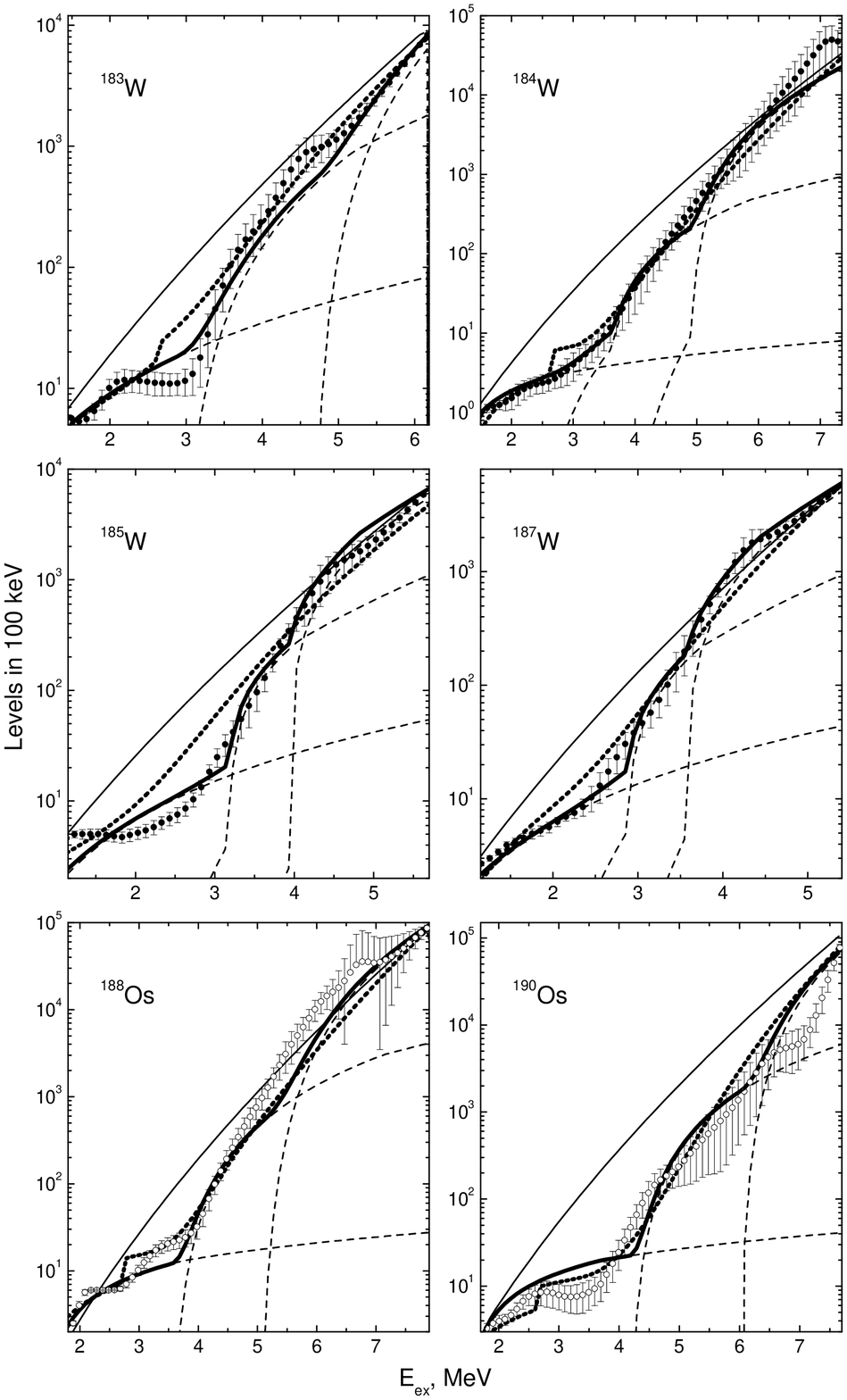}

{{\sl Fig. 7.}
 The same as in Fig. 2 for nuclei: $^{183,184,185,187}$W, $^{188,190}$Os.}\\
\end{figure}
\newpage
\begin{figure}
\vspace{2cm}
\leavevmode
\epsfxsize=14cm
\epsfbox{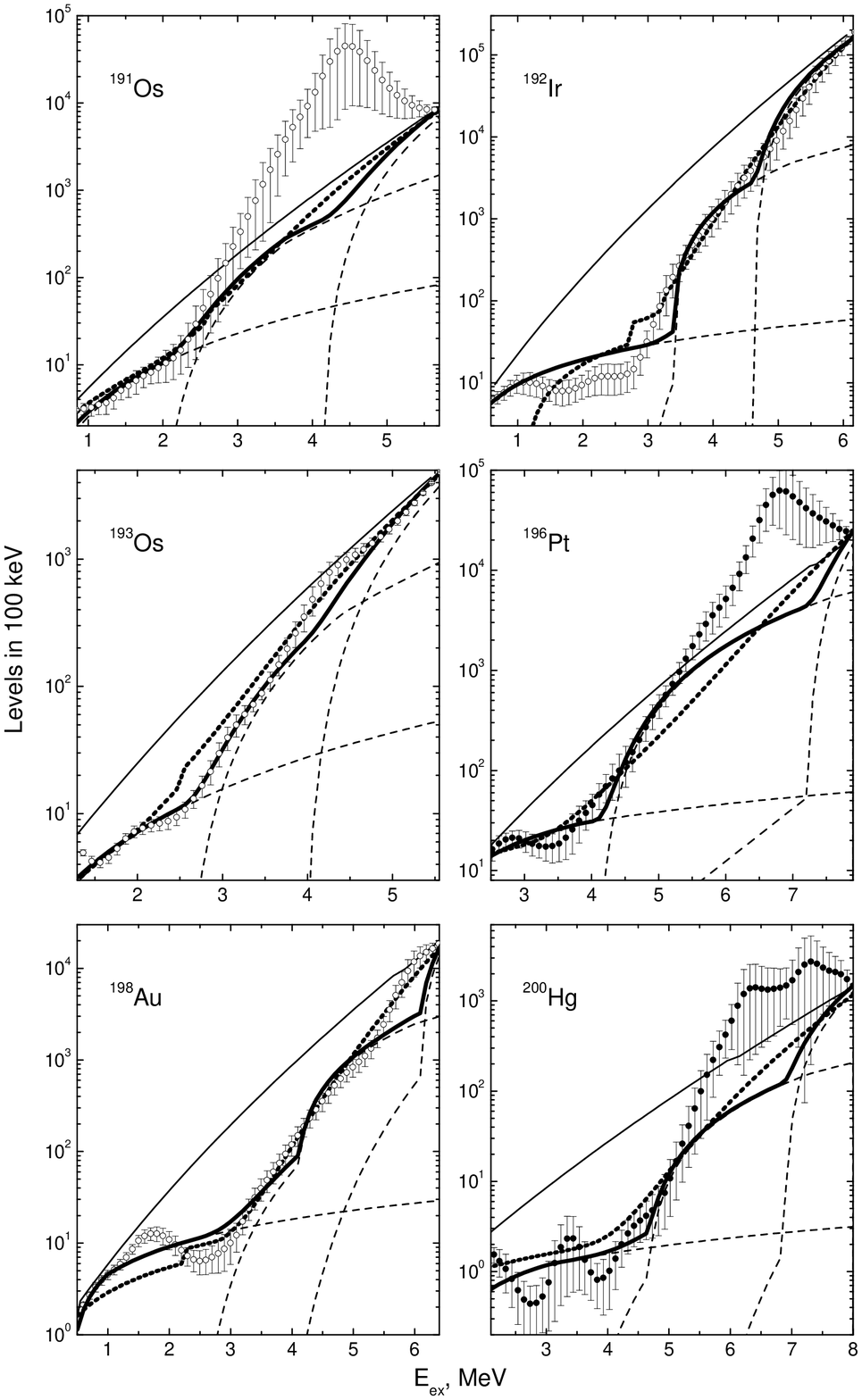}

{{\sl Fig. 8.} 
 The same as in Fig. 2 for nuclei: $^{191}$Os, $^{192}$Ir,
$^{193}$Os, $^{196}$Pt, $^{198}$Au, $^{200}$Hg.
}\\
\end{figure}
\end{document}